\documentclass[aps,prb,twocolumn,a4paper,superscriptaddress]{revtex4}
\usepackage{graphicx}
\usepackage{amsmath}

\usepackage[usenames,dvipsnames]{color}

\begin{document}
\title{A route towards finding large magnetic anisotropy in nano-composites: application to a W$_{1-x}$Re$_x$/Fe multilayer}

\author{Sumanta Bhandary}
\affiliation{Department of Physics and Astronomy, Uppsala University, Box 516,
 751\,20 Uppsala, Sweden}
 \author{Oscar Gr{\aa}n\"as}
\affiliation{Department of Physics and Astronomy, Uppsala University, Box 516,
 751\,20 Uppsala, Sweden}
 \author{Laszlo Szunyogh}
\affiliation{Department of Theoretical Physics, Budapest University of
Technology and Economics, Budafoki u. 8, H-1111, Budapest, Hungary}
\author{Biplab Sanyal}
\affiliation{Department of Physics and Astronomy, Uppsala University, Box 516,
 751\,20 Uppsala, Sweden}
 \author{Lars Nordstr\"om}
\affiliation{Department of Physics and Astronomy, Uppsala University, Box 516,
 751\,20 Uppsala, Sweden}
\author{Olle Eriksson}
\affiliation{Department of Physics and Astronomy, Uppsala University, Box 516,
 751\,20 Uppsala, Sweden}

\date{\today }

\begin{abstract}
We suggest here a novel nano-laminate, 5[Fe]/2[W$_x$Re$_{1-x}$] (x=0.6-0.8), with enhanced magnetic hardness in combination with a large saturation moment. The calculated magnetic anisotropy of this material reaches values of 5.3-7.0 MJ/m$^3$, depending on alloying conditions. We also propose a recipe in how to identify other novel magnetic materials, such as nano-laminates and multilayers, with large magnetic anisotropy in combination with a high saturation moment.
\end{abstract}
\maketitle

\section{Introduction}
The search for new materials with tailored properties for specific applications is a vital effort in the realization of novel and improved technologies. Recent developments in the ability to perform theoretical simulations with predictive power are just starting to show potential in this regard. An example here is the predicted huge tunneling magneto-resistance (TMR) of Fe/MgO/Fe trilayers,\cite{buttler,mathon} which subsequently was verified experimentally.\cite{TMRexpt} Most read-heads in modern disc-drives actually make use of this huge TMR effect. 

Magnetic materials are found in many applications. An example are permanent magnets, where the large energy product of Nd$_2$Fe$_{14}$B is utilized,\cite{ndfeb} e.g. in electrical machines and generators. The materials properties which are relevant in  Nd$_2$Fe$_{14}$B is the large magnetic anisotropy energy (MAE) combined with a large saturation moment (M$_s$). Recently the risk of only a limited access to rare-earth elements has been discussed, with a potential threat for manufacturing permanent magnets like Nd$_2$Fe$_{14}$B.\cite{physicstoday} Other classes of materials, which do not contain rare-earth elements, but with similar magnetic properties, are hence in high demand.

To identify materials with a large saturation moment only, is a subject of interest in its own right, with several challenges, e.g. as discussed recently.\cite{heiko} The wish of combining a large saturation moment with a high MAE becomes even more complicated, since the physical interactions responsible for one property do not necessarily correlate with those determining the other. To date no material has been identified with magnetic properties superior to rare-earth based compounds, where SmCo$_5$ and Nd$_2$Fe$_{14}$B stand out in particular. Several new materials have however been considered, where the tetragonal compound FePt is one of the most recently studied materials.\cite{FePt}

Theory based on density functional theory has proven accurate in reproducing measured saturation moments, and the difference between observations and measurements is seldom larger than 5-10 \%. As regards the magnetic anisotropy it is significantly more difficult, since the accuracy needed is very high. However, some developments have recently been made.\cite{laszlo} As an example we mention the predicted large MAE of a tetragonally strained FeCo-alloy\cite{burkert} which subsequently was verified experimemtally.\cite{gabi} 

In this communication we use first principles theory and focus on the magnetic anisotropy energy (MAE) and
suggest a novel nano-laminate with enhanced magnetic hardness in combination with a large saturation moment. In addition to identifying this new material, we propose here a recipe in how to identify other novel magnetic materials, such as nano-laminates and multilayers, with large magnetic anisotropy in combination with a high saturation moment.

\section{Geometry and Computational details}

In a similar spirit as that used in Ref.\onlinecite{burkert} we wish to investigate the MAE as a function of electron concentration, but in the present study it is the electron concentration of the ligand atoms, W and Re, which are under focus, not the atoms carrying the magnetic moment(Fe). The advantage of tuning the electron concentration of the ligand states, is that they can be chosen with a large spin-orbit coupling, which via hybridization with the Fe 3d states are expected\cite{candersson} to influence the anisotropy of the whole multilayer. We considered multilayers of different thickness of the Fe and W-Re layer, and the chosen composition, 5[Fe]/2[W$_x$Re$_{1-x}$] (x=0.6-0.8), is the one that gives a suitable strain of the W-Re layer, as well as an optimized electron concentration, as described below. The final geometry was obtained via force minimization using the VASP package. This resulted in essentially a tetragonal structure with an effective c/a ratio of the W-Re layers (and of the Fe layers), i.e. the ratio between atomic distances of the W-Re layer in the out-of-plane and in-plane direction. As we shall see below this ratio is a crucial parameter determining the MAE. 

Using the geometry optimized structure we then used two all electron methods to calculate the MAE, which are described in detail below. The two methods are below referred to as FPLMTO-VCA\cite{RSPt} and KKR-CPA.\cite{SKKRbook} In these calculations we also varied, within limits, the c/a ratio of the W-Re layer, and as we shall see below, this causes drastic changes in the MAE. 
An experimental realization of modifying the effective c/a ratio of multilayers is, in general, readily done by varying the thickness of the individual layers. 

As mentioned, from the geometry optimized structure we used an all electron, full-potential linear muffin-tin orbital (FPLMTO) method\cite{RSPt} to evaluate the MAE. This calculation used the simple virtual crystal approximation (VCA) to treat the alloying between W and Re. In these subsequent calculations we also varied, within limits, the c/a ratio of the W-Re layer, and as we shall see below, this causes drastic changes in the MAE. An experimental realization of modifying the effective c/a ratio of the W-Re layers can be done by tuning the relative amount of Fe and W-Re layers, i.e. by considering the superlattice m[Fe]/n[W$_x$Re$_{1-x}$], with m $\neq$ 5 and n $\neq$ 2.

In order to give further credit to our theoretical study we repeated the MAE
calculations for the 5[Fe]/2[W$_x$Re$_{1-x}$] superlattice
in terms of the fully relativistic screened Korringa-Kohn-Rostoker
(SKKR) method \cite{SKKRbook}.
This method can combine a full relativistic description with the more advanced coherent-potential approximation (CPA) to tackle the chemical disorder in the W$_x$Re$_{1-x}$ layers (where x=0.79). 
We used the local spin--density approximation of the density functional theory as parametrized by Vosko \emph{et al. }\cite{voskoCJP80}, the effective potentials and fields were treated within the atomic sphere approximation (ASA) with an angular momentum cut--off of $\ell_{max}=3$. 
To calculate the MAE we used
the so-called {\em magnetic force theorem}~\cite{jansenPRB99} in which the previously
determined self-consistent effective potentials and fields are kept fixed and
the change of total energy of the system with respect to the direction of
the magnetization, $\hat{n}$, is approached by that of
the single-particle (band) energy. In these calculations the energy integrations were performed by sampling 20 points on a semi-circular path in the upper complex semi-plane
and, for the energy point closest to the real axis,  nearly 10.000 $k$-points were 
selected in the surface Brillouin zone that ensured
a numerical error not exceeding about 5\% of the calculated MAE values.

\section{Results}

\begin{figure}[h]
\begin{center}
\includegraphics[scale=0.32]{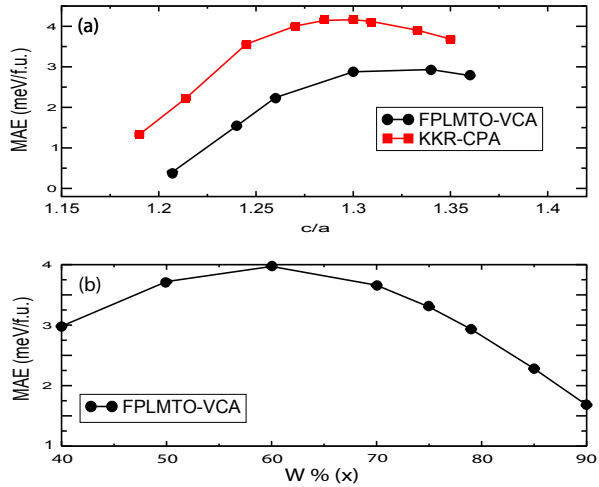}
\end{center}
\caption{(Color online) (a) Calculated magnetic anisotropy of 5[Fe]/2[W$_x$Re$_{1-x}$] (x=0.8) as a function of effective c/a ratio of the W-Re layers. (b) Calculated magnetic anisotropy of 5[Fe]/2[W$_x$Re$_{1-x}$] as a function of W concentration, $x$, for a c/a ratio of 1.34 of the W-Re layers. The MAE is shown per formula unit.} \label{fig1}
\end{figure}

We show in Fig.\ref{fig1}a the calculated MAE of 5[Fe]/2[W$_x$Re$_{1-x}$] as a function of the local c/a ratio of the W-Re layers. The starting concentration which was considered was x=0.8. In the figure we show both results from FPLMTO-VCA as well as KKR-CPA. From the figure it is clear that the two methods give similar overall behavior, but that the KKR-CPA calculations result in a maximum of $\sim$ 4 meV/f.u., whereas the FPLMTO-VCA results have a maximum of $\sim$ 3 meV/f.u. 
This is comforting since the KKR-CPA treats the disorder better while the FPLMTO-VCA handles the relaxed structure more accurately.
The two methods give both a similar shape of the MAE curve with respect to c/a ratio of the W-Re layer, and the maximum of the curves occur at similar c/a. The maximum value of the KKR-CPA results can be compared to the MAE of FePt, which from theory is $\sim$3 meV/f.u.,\cite{FePt2} whereas experiment results in 1.2 meV/f.u.\cite{FePt3} The MAE of 5[Fe]/2[W$_{0.8}$Re$_{0.2}$] is hence very comparable to that of FePt.
A further tuning of the magnetic anisotropy can be done by modifying the amount of alloying between W and Re in the multilayer, while keeping the c/a ratio the same as that giving the maximum MAE value in Fig.\ref{fig1}, i.e. 1.34. Calculations based on FPLMTO-VCA are shown in Fig.\ref{fig1}b, and we note that the MAE can be made $\sim$ 30 \% larger by tuning $x$ to 0.6.

As to the magnetic moments they are rather constant for the different values of x considered, and we obtain values of 1.65 $\mu_B$ per interface atom and 2.4 $\mu_B$ for the sub-interface Fe atoms, and 2.2 $\mu_B$ for the Fe atom in the middle of the Fe slab. The moment on W-Re is -0.1 $\mu_B$, resulting in a total moment just above 10 $\mu_B$/f.u. For comparison we note that for FePt this value is 3.3 $\mu_B$/f.u.

So far we have only considered calculated values per formula unit, whereas for practical applications it is with respect to volume one should compare the 5[Fe]/2[W$_x$Re$_{1-x}$] system with e.g. FePt. As concerns the MAE we obtain a maximum value 5.3-7.0 MJ/m$^3$ (depending on concentration or computational method) which should be compared to a measured value of $\sim$ 7 MJ/m$^3$ for FePt \cite{FePt}. For the moments we obtain 0.11 $\mu_B$/\AA$^{3}$ which is to be compared to 0.12 $\mu_B$/\AA$^{3}$ for FePt. A final note on the comparison to the properties of FePt, is that this material often exhibits partial chemical disorder on the Fe and Pt sublattices, which drastically reduces the MAE.\cite{staunton}

In order to obtain a more detailed, microscopical understanding for why the MAE of Fig.\ref{fig1} obtains a maximum at c/a close to 1.3 we show first of all that the large MAE is caused by the large spin-orbit coupling of the W-Re atoms. As a second step we will illustrate that it is a delicate detail of the energy bands of the W-Re layer, which take a special structure for a c/a ratio close to 1.3, which determine the large MAE. We start however, with analyzing the spin-orbit effect of the W-Re atoms. 

\begin{figure}[h]
\begin{center}
\includegraphics[scale=0.28]{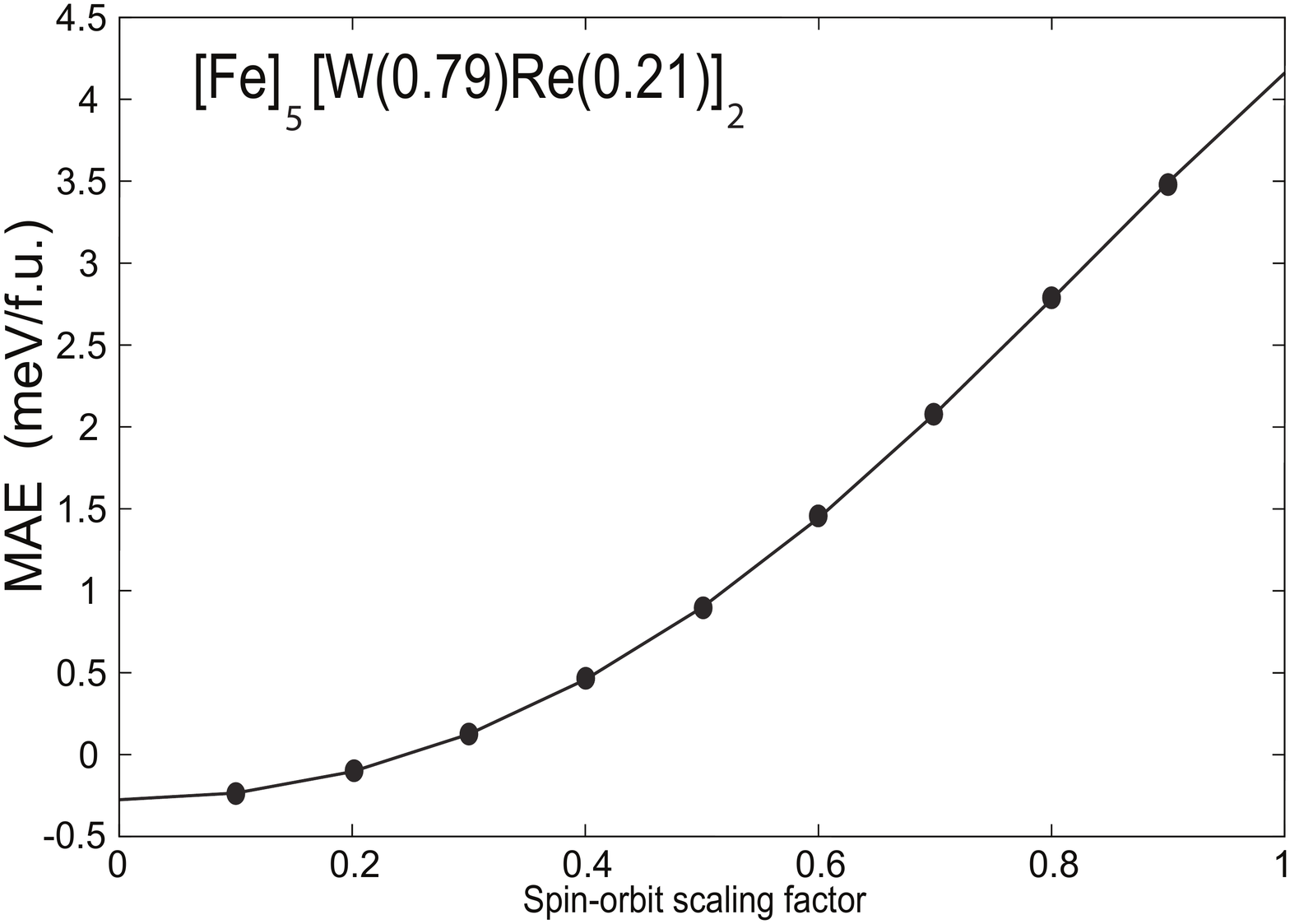}
\end{center}
\caption{Calculated MAE (circles) within the KKR-CPA approach with SOC strength scaled between
0 and 1.  The optimal setup, $c/a = 1.3$ and $x=0.8$, is fixed for the 5[Fe]~2[W$_x$Re$_{1-x}$] 
superlattice. Solid line serves as a guide for the eye.} \label{fig2}
\end{figure}

As a direct way to illustrate the dominant role of the W$_x$Re$_{1-x}$ layers 
in the formation of the high perpendicular MAE of this system, we varied the spin-orbit coupling strength at these layers for the optimal $c/a$ ratio ($c/a=1.3$) and
concentration (for $x=0.8$, other values of x give a similar picture). Here we employed the technique developed by Ebert {\em et al.} within the KKR formalism.~\cite{SOCscaling} The corresponding MAE values are shown in Fig.~\ref{fig2} for SOC strengths ($\xi$)
 scaled from $0$ to $1$. Remarkably, for $\xi=0$, i.e. when SOC is considered just in the Fe layers, the MAE
 is slightly negative preferring thus an in-plane orientation of the magnetization.  Increasing $\xi$ in the
  W$_x$Re$_{1-x}$ layers the MAE starts to increase as a quadratic function: this behavior is well
  understood in terms of the second order perturbation theory, discussed below.
  For about $0.5< \xi < 1$, however, the MAE seems to follow a linear behavior implying that 
in this regime SOC can not be regarded as a small second order perturbation to the scalar-relativistic 
spin-polarized band structure. Overall, the data of Fig.\ref{fig2} demonstrate that the large MAE is indeed caused by the large spin-orbit coupling of the W-Re atoms.

We now turn to investigating why an effective c/a ratio of the W-Re layer close to 1.3 causes a maximum value of the MAE. 
The MAE of a nano-composite material can be expressed as
\begin{equation}
\label{eq:deltaE}
\Delta E_{\scriptsize{SO}} = \sum_{qq'} \Delta E_{qq'} = \sum_{qq'}\left[E_{qq'}(\hat{n}_{1})-E_{qq'}(\hat{n}_{2})\right],
\end{equation}
where $q$ is the atomic species and $\hat{n}_{i}$ are two spin-quantization axis, one out of plane and one in plane. Each of the $E_{qq'}(\hat{n})$ can now, according to second order perturbation theory, be written as
\begin{equation}
\begin{split}
\label{eq:Eqss}
E_{qq'}(\hat{n})=-\sum_{\mathbf{k}ij} \sum_{\{s\}} \sum_{\{m\}} n_{\mathbf{k}is,qm,q'm'}n_{\mathbf{k}js',q'm'',qm''} \\
\times \frac{\langle qms | H_{\scriptsize{SO}}(\hat{n}) | qm'''s' \rangle \langle q'm''s' | H_{\scriptsize{SO}}(\hat{n}) | q'm's \rangle }{\epsilon_{\mathbf{k}j} - \epsilon_{\mathbf{k}i}},
\end{split}
\end{equation}
where $\mathbf{k}$ represents the sampling points in the Brillouin zone, $i$, $j$ is the occupied and unoccupied states,  $\{s\}=\{s,s'\}$ run over the spin components and $\{m\} = \{m, m', m'', m'''\}$ run over the magnetic quantum numbers. The basis functions $|qlms \rangle$ are specified through the atomic site $q$, the $l$, $m$ and $s$ quantum numbers, i.e.\  the azimuthal, magnetic and spin component quantum numbers. Note that the band character $n_{{\mathbf{k}is,qm,q'm'}}$ includes hybridization, i.e.\ mixing of different basis functions at e.g.\ different atomic sites $q$. The contributions from Eqn. \ref{eq:Eqss} are seen to become large if the energy difference between occupied and unoccupied electron states is small to permit a strong coupling of the spin-orbit interaction, especially if this happens for a large fraction of the BZ. This fact is further enhanced if the states that couple via the spin-orbit coupling have a significant contribution from the states residing on the heavy W-Re atom, since this guarantees that at least one of the spin-orbit matrix elements in Eqn.~\ref{eq:Eqss} is then large. 

\begin{figure}[h]
\begin{center}
\includegraphics[scale=0.32]{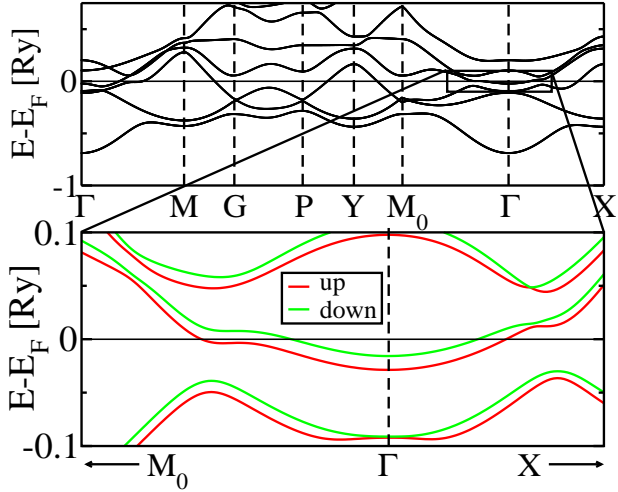}
\end{center}
\caption{(Color online) Non spin-polarized bands (upper panel) and spin-split bands (lower panel) of W-Re for bct c/a ratio 1.3.} \label{fig3}
\end{figure}

It is in such an analysis that the effective c/a ratio of W-Re close to 1.3 becomes clear. Inspection of a W-Re alloy reveals that as a function of c/a ratio, conspicuous features of the band structure explain the enhanced MAE. To illustrate this we show in Fig. \ref{fig3} the energy bands along $\Gamma-\mathrm{M}-\mathrm{G}-\mathrm{P}-\mathrm{Y}-\mathrm{M}_{0}-\Gamma-\mathrm{X}$ (of the bct BZ) of a W-Re alloy (calculated in this case by virtual crystal approximation - VCA - for simplicity, and with x=0.8). From the figure one may note a spin-degenerate band situated very close to the Fermi level ($E_\mathrm{F}$) between M$_{0}$ and $\Gamma$ and X. As a function of changing the c/a ratio this band moves from being completely occupied to become unoccupied, and for c/a=1.3 it is situated right at E$_F$. When put in contact to Fe, in the 5[Fe]/2[W$_x$Re$_{1-x}$] multilayer, the narrow band discussed in Fig.\ref{fig3} will become exchange split, due to the induced moment of the W-Re layers. In order to illustrate this effect in a simple way, we have performed 
a fixed spin moment calculation of W-Re, with the same induced moment on the W-Re site as for the 5[Fe]/2[W$_x$Re$_{1-x}$] multilayer. The resulting dispersion relationship is also shown in Fig.\ref{fig3} (bottom panel). Here we note a breaking of the spin-degeneracy close to $E_{F}$, and the flat band is split into up- and down-bands, which for a substantial part of the BZ are situated on each side of $E_{F}$ in a large part of the in-plane Brillouin zone. Spin-orbit coupling via the $l_{+}s_{-} + l_{-}s_{+}$ terms of $H_{SO}$ in Eqn.2 then provides a large contribution to the magnetic anisotropy. A modification of the c/a ratio away from this value moves the flat bands away from $E_{F}$, which reduces the MAE.

\section{Conclusion}
We have here presented theoretical evidence for that a non-laminate involving W-Re layers in contact to Fe, can result in a magnetic material with a substantial magnetic moment and a large magnetic anisotropy. The new material compares in performance to the well known material FePt, both in terms of saturation moment as well as anisotropy. One advantage with the present system is the possibility to tune the MAE by tuning the local c/a ratio of the W-Re layer, which can be done by introducing more Fe layers. This simultaneously also increases the saturation moment per volume.

\begin{figure}[h]
\begin{center}
\includegraphics[scale=0.25]{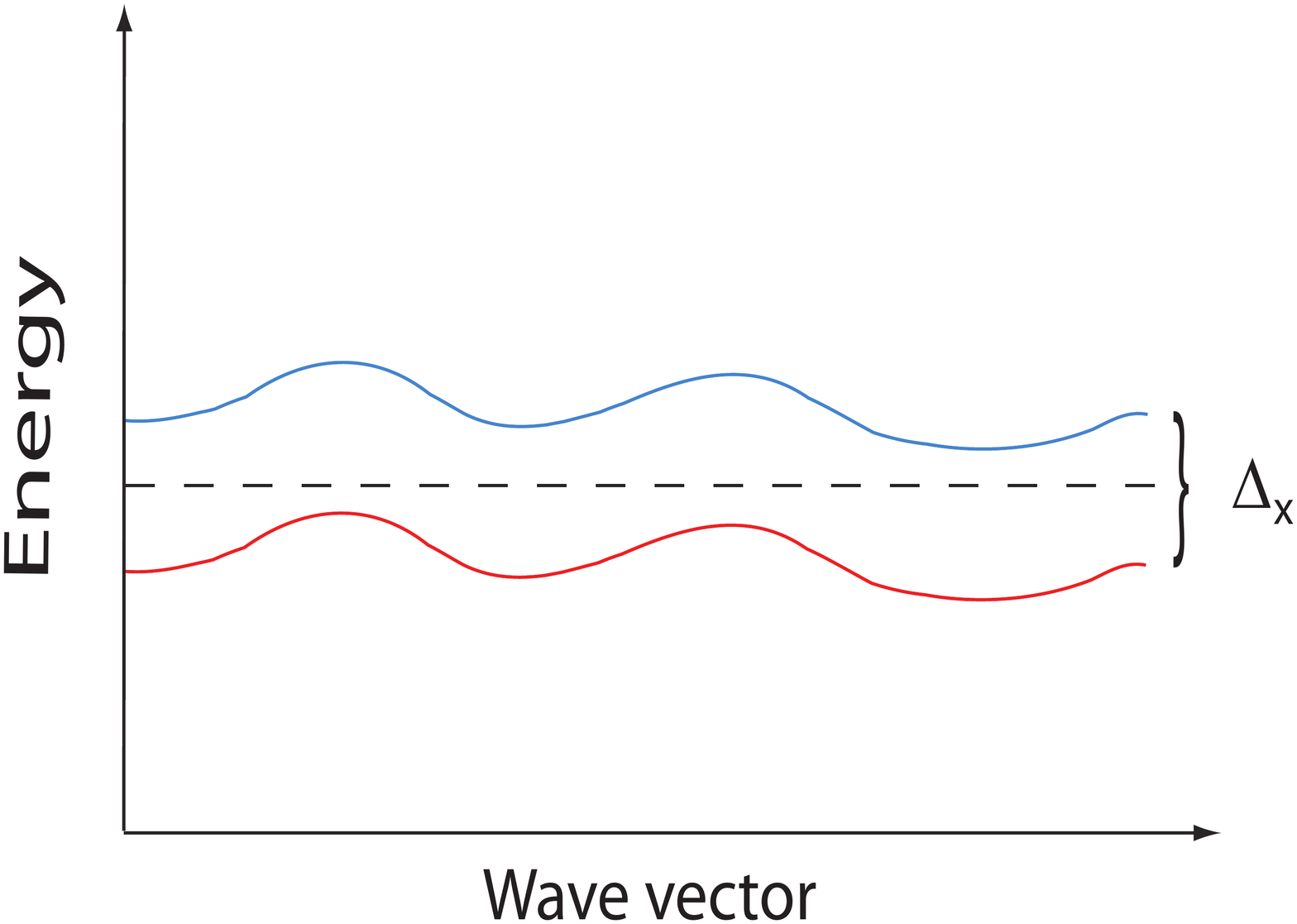}
\end{center}
\caption{(Color online) Schematic picture of the electronic structure of bands projected on heavy atoms, having possibility of inducing a large magnetic anisotropy. Majority spin states in red, minority states in blue, and the Fermi level a dashed horizontal line. The exchange splitting is marked as $\Delta_x$.} \label{fig4}
\end{figure}
 
Our analysis points to a general route to find materials with a large magnetic anisotropy. A strong ferromagnetic material, preferably based on Fe or an FeCo alloy, should be in contact with a heavier material with large spin-orbit splitting. Most crucially, the electronic structure of the heavy material should be similar to the schematic energy band plot of Fig.\ref{fig4}. Here bands which are dominated by states on the heavy atom are drawn to be exchange split, due to the interaction (hybridization) of the ferromagnetic material. If a large fraction of the BZ has majority bands below the Fermi level, and a large fraction of the BZ has minority states above (as Fig.\ref{fig4} is drawn), then the spin-orbit coupling of the heavy atom will give rise to a large contribution to the MAE for many wave vectors (as Eqn.2 shows). 

The situation shown in  Fig.\ref{fig4} corresponds to a set of narrow bands close to the Fermi level, which corresponds to a system with a large value of the density of states at the Fermi level (DOS($E_F$)). Hence, a large value  of DOS($E_F$) is also a requirement when searching for novel materials with a large MAE. Unfortunately a large value of DOS($E_F$) is often associated with an unstable or meta-stable crystal structure,\cite{skriver} which points to that it is meta-stable systems, such as multilayers and nano-laminates, which one should consider, when searching for novel materials with a large MAE.


\section{Acknowledgement}
We gratefully acknowledge financial support from the Swedish
Research Council (VR), Swedish Foundation for Strategic Research
(SSF), G\"oran Gustafssson Foundation, Carl Tryggers Foundation, STEM and STINT. O.E. is in addition grateful to the ERC (project 247062 - ASD) and KAW foundation for support. Support from eSSENCE and STANDUP acknowledged.
We also acknowledge Swedish National Infrastructure for Computing (SNIC)
for the allocation of time in high performance supercomputers.
Partial financial support was provided also by the Hungarian Research Foundation (contract no. OTKA K68312, K77771
and IN83114) and by the New Szechenyi Plan of Hungary (Project ID: T\'AMOP-4.2.1/B-09/1/KMR-2010-0002).



\begin{thebibliography}{25}
\expandafter\ifx\csname natexlab\endcsname\relax\def\natexlab#1{#1}\fi
\expandafter\ifx\csname bibnamefont\endcsname\relax
  \def\bibnamefont#1{#1}\fi
\expandafter\ifx\csname bibfnamefont\endcsname\relax
  \def\bibfnamefont#1{#1}\fi
\expandafter\ifx\csname citenamefont\endcsname\relax
  \def\citenamefont#1{#1}\fi
\expandafter\ifx\csname url\endcsname\relax
  \def\url#1{\texttt{#1}}\fi
\expandafter\ifx\csname urlprefix\endcsname\relax\def\urlprefix{URL }\fi
\providecommand{\bibinfo}[2]{#2}
\providecommand{\eprint}[2][]{\url{#2}}

\bibitem[{\citenamefont{Beaurepaire et~al.}(1996)\citenamefont{Beaurepaire,
  Merle, Daunois, and Bigot}}]{buttler}
\bibinfo{author}{\bibfnamefont{W.H.}~\bibnamefont{Butler}},
\bibinfo{author}{\bibfnamefont{X.G.}~\bibnamefont{Zhang}},
\bibinfo{author}{\bibfnamefont{T.C.}~\bibnamefont{Schultess}},
\bibinfo{author}{\bibfnamefont{J.M.}~\bibnamefont{MacLaren}},
  \bibinfo{journal}{Physical Review B} \textbf{\bibinfo{volume}{63}},
  \bibinfo{pages}{54416} (\bibinfo{year}{2001}).

  \bibitem[{\citenamefont{Beaurepaire et~al.}(1996)\citenamefont{Beaurepaire,
  Merle, Daunois, and Bigot}}]{mathon}
\bibinfo{author}{\bibfnamefont{J.}~\bibnamefont{Mathon}},
\bibinfo{author}{\bibfnamefont{A.}~\bibnamefont{Umarski}},
  \bibinfo{journal}{Physical Review B} \textbf{\bibinfo{volume}{63}},
  \bibinfo{pages}{220403} (\bibinfo{year}{2001}).



  \bibitem[{\citenamefont{Beaurepaire et~al.}(1996)\citenamefont{Beaurepaire,
  Merle, Daunois, and Bigot}}]{TMRexpt}
\bibinfo{author}{\bibfnamefont{S.}~\bibnamefont{Yuasa}},
\bibinfo{author}{\bibfnamefont{T.}~\bibnamefont{Nagahama}},
\bibinfo{author}{\bibfnamefont{A.}~\bibnamefont{Fukushima}},
\bibinfo{author}{\bibfnamefont{Y.}~\bibnamefont{Suzuki}},
\bibinfo{author}{\bibfnamefont{K.}~\bibnamefont{Ando}},
  \bibinfo{journal}{Nature Materials} \textbf{\bibinfo{volume}{3}},
  \bibinfo{pages}{868} (\bibinfo{year}{2004}).



   \bibitem[{\citenamefont{ndfeb}(1996)\citenamefont{ndfeb}}]{ndfeb}
\bibinfo{author}{\bibfnamefont{J.M.D.}~\bibnamefont{Coey}},
  \bibinfo{journal}{Rare-earth Iron Permanent Magnets} \textbf{\bibinfo{volume}{Oxford}},
  \bibinfo{pages}{~}(\bibinfo{year}{1996}).

    \bibitem[{\citenamefont{physicstoday}(1996)\citenamefont{physicstoday}}]{physicstoday}
\bibinfo{author}{\bibfnamefont{D.}~\bibnamefont{Kramer}},
  \bibinfo{journal}{Physics Today, May,} \bibinfo{pages}{22},
  \bibinfo{volume}{} (\bibinfo{year}{2010}).



  \bibitem[{\citenamefont{Beaurepaire et~al.}(1996)\citenamefont{Beaurepaire,
  Merle, Daunois, and Bigot}}]{heiko}
\bibinfo{author}{\bibfnamefont{B.}~\bibnamefont{Sanyal}},
\bibinfo{author}{\bibfnamefont{C.}~\bibnamefont{Antoniak}},
\bibinfo{author}{\bibfnamefont{B.}~\bibnamefont{Krumme}},
\bibinfo{author}{\bibfnamefont{A.}~\bibnamefont{Warland}},
\bibinfo{author}{\bibfnamefont{F.}~\bibnamefont{Stromberg}},
\bibinfo{author}{\bibfnamefont{K.}~\bibnamefont{Fauth}},
\bibinfo{author}{\bibfnamefont{H.}~\bibnamefont{Wende}},
\bibinfo{author}{\bibfnamefont{O.}~\bibnamefont{Eriksson}},
  \bibinfo{journal}{Physical Review Letters} \textbf{\bibinfo{volume}{104}},
  \bibinfo{pages}{156402} (\bibinfo{year}{2010}).
  

  \bibitem[{\citenamefont{Beaurepaire et~al.}(2000)\citenamefont{Beaurepaire,
  Merle, Daunois, and Bigot}}]{FePt}
\bibinfo{author}{\bibfnamefont{S.}~\bibnamefont{Sun}},
\bibinfo{author}{\bibfnamefont{C.B.}~\bibnamefont{Murray}},
\bibinfo{author}{\bibfnamefont{D.}~\bibnamefont{Weller et al.}},
  \bibinfo{journal}{Science} \textbf{\bibinfo{volume}{287}},
  \bibinfo{pages}{1989} (\bibinfo{year}{2000}).

  \bibitem[{\citenamefont{Beaurepaire et~al.}(1996)\citenamefont{Beaurepaire,
  Merle, Daunois, and Bigot}}]{laszlo}
\bibinfo{author}{\bibfnamefont{L.}~\bibnamefont{Szunyogh}},
\bibinfo{author}{\bibfnamefont{B.}~\bibnamefont{Ujfalussy}},
\bibinfo{author}{\bibfnamefont{P.}~\bibnamefont{Weinberger}},
  \bibinfo{journal}{Physical Review B} \textbf{\bibinfo{volume}{51}},
  \bibinfo{pages}{9552} (\bibinfo{year}{1995}).


  \bibitem[{\citenamefont{Andersson et~al.}(2006)\citenamefont{Andersson, Burkert, Warnicke, Bj\"ork, Sanyal,
Chacon, Zloeta, Nordstr\"om, Nordblad and Eriksson}}]{burkert}
\bibinfo{author}{\bibfnamefont{T.}~\bibnamefont{Burkert}},
\bibinfo{author}{\bibfnamefont{L.}~\bibnamefont{Nordstr\"om}},
\bibinfo{author}{\bibfnamefont{O.}~\bibnamefont{Eriksson} ~\bibnamefont{and}}
\bibinfo{author}{\bibfnamefont{O.}~\bibnamefont{Heinonen}},
  \bibinfo{journal}{Physical Review Letters} \textbf{\bibinfo{volume}{93}},
  \bibinfo{pages}{27203} (\bibinfo{year}{2004}).

  \bibitem[{\citenamefont{Andersson et~al.}(2006)\citenamefont{Andersson, Burkert, Warnicke, Bj\"ork, Sanyal,
Chacon, Zloeta, Nordstr\"om, Nordblad and Eriksson}}]{gabi}
\bibinfo{author}{\bibfnamefont{G.}~\bibnamefont{Andersson}},
\bibinfo{author}{\bibfnamefont{T.}~\bibnamefont{Burkert}},
\bibinfo{author}{\bibfnamefont{P.}~\bibnamefont{Warnicke}},
\bibinfo{author}{\bibfnamefont{M.}~\bibnamefont{Bj\"ork}},
\bibinfo{author}{\bibfnamefont{B.}~\bibnamefont{Sanyal}},
\bibinfo{author}{\bibfnamefont{C.}~\bibnamefont{Chacon}},
\bibinfo{author}{\bibfnamefont{C.}~\bibnamefont{Zloeta}},
\bibinfo{author}{\bibfnamefont{L.}~\bibnamefont{Nordstr\"om}},
\bibinfo{author}{\bibfnamefont{P.}~\bibnamefont{Nordblad}~\bibnamefont{and} }
\bibinfo{author}{\bibfnamefont{O.}~\bibnamefont{Eriksson}},
  \bibinfo{journal}{Physical Review Letters} \textbf{\bibinfo{volume}{96}},
  \bibinfo{pages}{37205} (\bibinfo{year}{2006}).
    
  
   \bibitem[{\citenamefont{Andersson et~al.}(2006)\citenamefont{C.Andersson}}]{candersson}
\bibinfo{author}{\bibfnamefont{C.}~\bibnamefont{Andersson}},
\bibinfo{author}{\bibfnamefont{B.}~\bibnamefont{Sanyal} }
\bibinfo{author}{\bibfnamefont{O.}~\bibnamefont{Eriksson et al.}},
  \bibinfo{journal}{Physical Review Letters} \textbf{\bibinfo{volume}{99}},
  \bibinfo{pages}{177207} (\bibinfo{year}{2007}).
  
\bibitem{RSPt}
J.M.Wills, M.Alouani, P.Andersson, A.Delin, O.Eriksson, A.Grechnev, 
``Full-Potential Electronic Structure Method, Energy and Force Calculations with Density Functional and Dynamical Mean Field Theory'' (Springer Series in Solid-State Sciences, Volume 167, 2010).

\bibitem{SKKRbook}
J. Zabloudil, R. Hammerling, L. Szunyogh, and P. Weinberger,
Electron Scattering in Solid Matter, Springer Series in Solid-
State Sciences Vol. 147 (Springer, Heidelberg, 2005)

\bibitem{voskoCJP80}
S.~H. Vosko, L. Wilk, and M. Nusair, Can. J. Phys. {\bf 58},  1200  (1980).

\bibitem{jansenPRB99}
H.~J.~F. Jansen, Phys. Rev. B {\bf 59},  4699  (1999).



\bibitem{FePt2}
G. H. O. Daalderop, P. J. Kelly, and M. F. H. Schuurmans,
Phys. Rev. B 44, 12054 (1991).

\bibitem{FePt3}
S. Okamoto, N. Kikuchi, O. Kitakami, T. Miyazaki, Y. Shimada,
and K. Fukamichi, Phys. Rev. B 66, 024413 (2002).

\bibitem{staunton} J.B. Staunton, S. Ostanin, S.S.A. Razee, B.L. Gyorffy, L. Szunyogh, B. Ginatempo, and E. Bruno, J. Phys. Cond. Matt. {\bf 16}, S5623 (2004). 

\bibitem{SOCscaling}
H. Ebert, H. Freyer and M. Deng, Phys. Rev. B {\bf 56}, 9454 (1997). 


\bibitem{skriver}
H.L.Skriver, Phys. Rev. B {\bf 31}, 1909 (1985). 

\end{thebibliography}
\end{document}